\def\Statusstring{}
\documentclass[12pt,letter]{article}

\usepackage{amsmath, amsthm, amssymb}
\usepackage{amssymb,epsfig}
\usepackage{pstricks} 

\def\qed{\hskip 3pt \hbox{\vrule width4pt depth2pt height6pt}}

\newtheorem{Lemma}{Lemma}

\newtheorem{Theorem}[Lemma]{Theorem}
\newtheorem{Proposition}[Lemma]{Proposition}
\newtheorem{Corollary}[Lemma]{Corollary}

\newcommand{\Aut}{\mathop{\mathrm{Aut}}\nolimits}

\newcommand{\Cay}{\mathop{\mathrm{Cay}}\nolimits}

\newcommand{\bZ}{\mathop{\mathrm{\mathbb{Z}}}\nolimits}

\begin{document}

\title{Automorphism groups of Cayley graphs generated by connected transposition sets}
\author{Ashwin Ganesan%
  \thanks{Department of Mathematics, Amrita School of Engineering, Amrita University, Amritanagar, Coimbatore - 641~112, Tamil Nadu, India.
  Email: \texttt{ashwin.ganesan@gmail.com, g\_ashwin@cb.amrita.edu}. }}
\date{}

\maketitle

\vspace{-6.5cm}
\begin{flushright}
  \texttt{\Statusstring}\\[1cm]
\end{flushright}
\vspace{+4.3cm}

\begin{abstract} 
Let $S$ be a set of transpositions that generates the symmetric group $S_n$, where $n \ge 3$. The transposition graph $T(S)$ is defined to be the graph with vertex set $\{1,\ldots,n\}$ and with vertices $i$ and $j$ being adjacent in $T(S)$ whenever $(i,j) \in S$.  
We prove that if the girth of the transposition graph $T(S)$ is at least 5, then the automorphism group of the Cayley graph $\Cay(S_n,S)$ is the semidirect product $R(S_n) \rtimes \Aut(S_n,S)$, where $\Aut(S_n,S)$ is the set of automorphisms of $S_n$ that fixes $S$.  This  
strengthens  a result of Feng on transposition graphs that are trees.
We also prove that if the transposition graph $T(S)$ is a $4$-cycle, then 
the set of automorphisms of the Cayley graph $\Cay(S_4,S)$ that fixes a vertex and each of its neighbors is isomorphic to the Klein 4-group and hence is nontrivial. 
We thus identify the existence of 4-cycles in the transposition graph as being an important factor in causing a potentially larger automorphism group of the Cayley graph.

\end{abstract}

\bigskip
\noindent\textbf{Index terms} --- Cayley graphs; transposition sets; automorphisms of graphs;
\\modified bubble-sort graph.
\bigskip


\section{Introduction}

Let $\Gamma=(V,E)$ be a simple, undirected graph. 
The set of all automorphisms of $\Gamma$ forms a permutation group called the automorphism group of $\Gamma$, which we denote by $\Aut(\Gamma)$.  
Given a group $H$ and a subset $S \subseteq H$, the Cayley digraph 
of $H$ with respect to $S$, denoted by $\Cay(H,S)$, is the digraph with vertex set $H$ and with an arc from $h$ to $sh$ whenever $h \in H$ and $s \in S$.  When $S$ is closed under inverses, 
$(g,h)$ is an arc of the Cayley digraph if and only if $(h,g)$ is an arc, and so we can identify the two arcs $(g,h)$ and $(h,g)$ with the undirected edge $\{g,h\}$.  When $1 \notin S$, the Cayley digraph contains no self-loops.  Thus, when $1 \notin S=S^{-1}$, $\Cay(H,S)$ can be considered to be a simple, undirected graph. The Cayley graph $\Cay(H,S)$ is connected if and only if $S$ generates $H$.  

The automorphism group of the Cayley graph $\Cay(H,S)$ contains the right regular representation $R(H)$ as a subgroup, and hence all Cayley graphs are vertex-transitive. Let $e$ denote the identity element of the group $H$ and also the corresponding vertex of $\Cay(H,S)$.  Since $R(H)$ is regular, $\Aut(\Gamma) = \Aut(\Gamma)_e R(H)$, where $\Aut(\Gamma)_e$ is the stabilizer of $e$ in $\Aut(\Gamma)$.  The set of automorphisms of the group $H$ that fixes $S$ setwise, denoted by $\Aut(H,S):= \{ \pi \in \Aut(H): S^{\pi}=S\}$, is a subgroup of $\Aut(\Gamma)_e$ (cf. \cite{Biggs:1993}).  For any Cayley graph $\Gamma=\Cay(H,S)$, the normalizer $N_{\Aut(\Gamma)} (R(H))$ is equal to the semidirect product $R(H) \rtimes \Aut(H,S)$ (cf. \cite{Godsil:1981}, \cite{Xu:1998}).   A Cayley graph $\Gamma:=\Cay(H,S)$ is said to be normal if $R(H)$ is a normal subgroup of $\Aut(\Gamma)$, or equivalently, if $\Aut(\Gamma) = R(H) \rtimes \Aut(H,S)$.  Thus, normal Cayley graphs are those that have the smallest possible full
automorphism group (cf. \cite{Xu:1998}).  An open problem in the literature is to determine which Cayley graphs are normal.   While one can often obtain some automorphisms of a graph, it is often difficult to prove that one has obtained the (full) automorphism group. 

Let $S$ be a set of transpositions in the symmetric group $S_n$. The transposition graph $T(S)$ is defined to be the graph with vertex set $\{1,2,\ldots,n\}$ and with vertices $i$ and $j$ being adjacent in $T(S)$ whenever $(i,j) \in S$.  A set of transpositions of $\{1,\ldots,n\}$ generates $S_n$ if and only if the transposition graph $T(S)$ is connected; thus, a set of transpositions is a minimal generating set for $S_n$ if and only if $T(S)$ is a tree (cf. \cite{Godsil:Royle:2001}).  In the sequel, we assume that the set of transpositions $S$ generates $S_n$; equivalently, we assume the graph $T(S)$ is connected.

Let $n \ge 3$.  Feng \cite{Feng:2006} showed that if $S$ is a set of transpositions in $S_n$, then $\Aut(S_n,S) \cong \Aut(T(S))$.  This result holds even if $T(S)$ is not a tree.  In the special case when $T(S)$ is a tree, Feng \cite{Feng:2006} showed that the automorphism group of the Cayley graph $\Cay(S_n,S)$ is the semidirect product $R(S_n) \rtimes \Aut(S_n,S)$.  

In the present paper, we generalize and extend the results of Feng on trees mentioned in the previous graph to arbitrary transposition graphs that may contain cycles.  The girth of a graph is the length of the shortest cycle in the graph.  In particular, trees have infinite girth.  We prove that if the girth of the transposition graph is at least 5, then the set of automorphisms of the Cayley graph $\Cay(S_n,S)$ that fixes a vertex and each of its neighbors is trivial and the automorphism group of the Cayley graph $\Cay(S_n,S)$ is the semidirect product $R(S_n) \rtimes \Aut(S_n,S)$.  On the other hand, if $T(S)$ is a 4-cycle, then the set of automorphisms of $\Cay(S_4,S)$ that fixes a vertex and each of its neighbors is shown to be the Klein 4-group and hence is not trivial.  These results thus identify 4-cycles in the transposition graph as playing an important role in leading to a potentially larger automorphism group (of the Cayley graph) than otherwise.

\bigskip The main result of this paper is the following:

\begin{Theorem} \label{theorem:main}
Let $S$ be a set of transpositions that generates $S_n$, where $n \ge 3$.  If the girth of the transposition graph $T(S)$ is at least 5, then the automorphism group of the Cayley graph $Cay(S_n,S)$ is the semidirect product $R(S_n) \rtimes \Aut(S_n,S)$, where $\Aut(S_n,S)$ is the set of automorphisms of $S_n$ that fixes $S$ setwise.  
\end{Theorem}

\bigskip \textbf{Remark}: A special case of Theorem~\ref{theorem:main} is the result of Feng \cite{Feng:2006} that if the transposition graph $T(S)$ is a tree, then the automorphism group of the Cayley graph $\Cay(S_n,S)$ is the semidirect product $R(S_n) \rtimes \Aut(S_n,S)$.  Further special cases of this result are the following: the result of Godsil and Royle \cite{Godsil:Royle:2001} that if $T(S)$ is an asymmetric tree, then $\Cay(S_n,S)$ has automorphism group $R(S_n) \cong S_n$; the result of Huang and Zhang \cite{Huang:Zhang:STn:submitted} that if $T(S)$ is the star $K_{1,n-1}$, then $\Cay(S_n,S)$ has automorphism group isomorphic to $S_n S_{n-1}$; and the result of Zhang and Huang \cite{Zhang:Huang:2005} that if $T(S)$ is the path graph then $\Cay(S_n,S)$ has automorphism group isomorphic to $S_n \mathbb{Z}_2$ .

\section{Proof of Theorem~\ref{theorem:main}}

We assume throughout that $S$ is a set of transpositions of $\{1,\ldots,n\}$ that generates $S_n$.  Thus the transposition graph $T(S)$ is connected.

When $n=3$ and $T(S)$ is the 3-cycle graph, the Cayley graph $\Cay(S_n,S)$ is the complete bipartite graph $K_{3,3}$ (cf. Biggs \cite[Chapter 16]{Biggs:1993}).  The complement of $K_{3,3}$ is the union of two disjoint 3-cycles.  Thus, the automorphism group of $K_{3,3}$ is the automorphism group of a partition of a six element set into two equal sized subsets and hence is equal to the semidirect product $(S_3 \times S_3) \rtimes \mathbb{Z}_2$ (cf. \cite[p. 46]{Dixon:Mortimer:1993}).

Thus, the cases where $n \le 3$ are easily dealt with; we assume in the sequel that $n \ge 4$. We first recall a preliminary result.


\begin{Lemma} (Godsil and Royle \cite[Lemma 3.10.3]{Godsil:Royle:2001}) \label{GodsilRoyle:4cycles}
Let $S$ be a set of transpositions such that the transposition graph $T(S)$ does not contain triangles, and let $t,k \in S$. Then, $tk = kt$ if and only if there is a unique 4-cycle in $\Cay(S_n,S)$ containing $e, t$ and $k$.
\end{Lemma}

\begin{Lemma} (Feng  \cite[Corollary 2.5]{Feng:2006}) \label{lemma:feng:normalTS}
Let $S$ be a set of transpositions generating $S_n$ satisfying the following condition for any two distinct transpositions $t,k \in S$: $tk = kt$ if and only if $\Cay(S_n,S)$ has a unique 4-cycle containing $e,t$ and $k$, and if $tk \ne kt$ then $\Cay(S_n,S)$ has a unique 6-cycle containing $e,t,k$ and a vertex at distance 3 from $e$.  Then $\Cay(S_n,S)$ is normal and has automorphism group isomorphic to $R(S_n) \rtimes \Aut(S_n,S)$.
\end{Lemma}

\begin{Theorem} \label{thm:6cycle:TS}
Let $S$ be a set of transpositions generating $S_n$, and let $\Gamma$ be the Cayley graph $\Cay(S_n,S)$.  Let $k, t \in S$ be distinct transpositions.  If $tk \ne kt$ and the girth of the transposition graph $T(S)$ is at least 5, then there is a unique 6-cycle in $\Gamma$ containing $e,t,k$ and a vertex at distance 3 from $e$.  
\end{Theorem}

\noindent \emph{Proof: } Suppose $T(S)$ does not contain triangles or 4-cycles, and let $t,k \in S$ be distinct transpositions such that $tk \ne kt$.  We show that there is a unique 6-cycle in $\Gamma$ containing $e, t, k$ and a vertex at distance 3 from $e$.  Since $tk \ne kt$, $t$ and $k$ are adjacent edges in $T(S)$, so we may assume without loss of generality that $k=(1,2)$ and $t=(2,3)$.  Then, $(e, t, kt, tkt=ktk, tk, k, e)$ is a 6-cycle in $\Gamma$.  The distance from $e$ to $tkt$ is at most 3. Also, $\Gamma$ is a bipartite graph with bipartition equal to the even and odd permutations, and so the distance from $e$ to $tkt$ is odd. But this distance cannot be 1 since $tkt = (1,3) \notin S$ because $T(S)$ has no triangles.  Thus, $tkt$ has distance 3 to $e$.  We show next that this 6-cycle is the unique 6-cycle in $\Gamma$ containing $e, t, k$ and a vertex at distance 3 from $e$.

Suppose $(e, t, t_1 t, t_2 t_1 t = k_2 k_1 k, k_1 k, k,e)$ is a 6-cycle in $\Gamma$ containing $e, t, k$ and a vertex at distance 3 from $e$.  Since the 6 vertices in this 6-cycle are distinct, we have that $t_1 \ne t, t_2 \ne t_1, k_2 \ne t_2, k_2 \ne k_1$ and $k_1 \ne k$. We consider 3 cases, depending on whether $k_1=t$ and $t_1=k$.

Case 1:  Suppose $k_1=t$ and $t_1=k$.   Then $k_2 k_1 k = t_2 t_1 t$ becomes $k_2 tk=t_2 k t$, or $t_2 k_2 = kt kt = (1,2,3)$.  Since $T(S)$ has no triangles, the only way to express $(1,2,3)=tk$ as a product of two transpositions $t_2 k_2$ is as $t_2 = t, k_2=k$.  This gives rise to the same 6-cycle given initially, and hence there exists only 6-cycle in $\Gamma$ containing $e,t,k$ and a vertex at distance 3 from $e$.

Case 2: Assume $k_1 \ne t$.  This means $k_1 \ne (2,3)$. 
We know $k_1 \ne k=(1,2)$ and since $T(S)$ has no triangles, $k_1 \ne (1,3)$. Thus, $\{1,2,3\}$ and the support of $k_1$ have an intersection of size at most 1. (For example, if $k_1=(1,4)$ say, then the intersection is $\{1\}$, and if $k_1=(4,5)$ say, then the intersection is trivial.)  This implies that the union of $\{1,2,3\}$ and the support of $k_1$ is at least 4 because $k_1$ contributes at least one new element to this union.  By $t_2 t_1 t = k_2 k_1 k$, we have $k_2 t_2 t_1 = k_1 k t$.  We now prove that the assumption that $T(S)$ have girth at least 5 forces $\{k_2, t_2, t_1\} = \{k_1, k, t\}$.  Recall that $k_1 \ne (2,3)$ since $T(S)$ does not contain
triangles, and $k$ and $t$ have total support $\{1,2,3\}$.

Suppose $k_1$ has overlapping support with $\{1,2,3\}$.  This can happen if and only if the three edges $\{k_1,k,t\}$ form a tree spanning 4 distinct vertices of $T(S)$, namely the vertices $\{1,2,3,i\}$ for some $i$. The product $k_1 k t$ of three transpositions that form a subtree in $T(S)$ is a permutation that consists of a single four-cycle, and conversely (cf. Godsil and Royle \cite[Lemma 3.10.2]{Godsil:Royle:2001}), if the product of three transpositions is a four-cycle, then these 3 transpositions constitute the edges of a spanning tree on the same 4 vertices corresponding to the support of the four-cycle.  Thus, the edges $\{k_2, t_2, t_1\}$ also form a spanning tree on the same 4 vertices $\{1,2,3,i\}$.  Now, the only spanning tree subgraph in $T(S)$ on these 4 vertices is the spanning tree $\{k_1,k,t\}$ because $T(S)$ does not contain triangles or 4-cycles.  
that contains a triangle or a 4-cycle.)
Thus, $\{k_2, t_2, t_1\} = \{k_1,k,t\}$, as claimed.

If $k_1$ has disjoint support from $\{1,2,3\}$, then the product $k_1 k t$ has a 2-cycle, say $(i,j)$ and a 3-cycle $kt = (1,3,2)$, the two cycles being disjoint.  By $k_2 t_2 t_1 = k_1 kt$, we have that at least one of $k_2, t_2$ or $t_1$ must equal $(i,j)$, and the product of the other two must equal $(1,3,2)$.  Since $T(S)$ has no triangles, the only way to decompose $(1,3,2)$ as a product of two transpositions from $S$ is as $(1,2)(2,3)=kt$.  Thus, two of $k_2,t_2,t_1$ must equal $k$ and $t$, and so we again get $\{k_2, t_2, t_1\} = \{k_1,k,t\}$.

Thus, if the girth of $T(S)$ is at least 5 and $k_1 \ne (2,3)$, then the union of the support of $k_1,k$ and $t$ is at least 4, and hence $k_2 t_2 t_1 = k_1 k t$ forces $\{k_2,t_2,t_1\}=\{k_1,k,t\}$.   A few subcases now arise, each of which is seen to be impossible:

We already have $k_2 \ne k_1$.  We consider the two cases $k_2 = k$ and $k_2 =t$:

Subcase 2.1:  $k_2=k$.  Then, either $t_2=k_1$ and $t_1=t$ or $t_2=t$ and $t_1=k_1$.  The first subcase is impossible because $t_1 \ne t$. The second subcase $t_2=t$ and $t_1=k_1$, along with $k_2 t_2 t_1 = k_1 k t$ gives $ktk_1 = k_1 kt$, i.e. $k_1 = (1,2,3) k_1 (1,3,2)$.  This equation implies that the support of $k_1$ is disjoint from $\{1,2,3\}$, and so $k_2 k_1 k = k_2 k k_1 = k_1 \in S$, implying that the vertex $k_2 k_1 k$ is at distance 1 and not distance 3 from $e$, a contradiction.

Subcase 2.2: $k_2 = t$:  Then we have two subcases - that either $t_2 = k_1$ and $t_1=k$ or $t_2=k$ and $t_1=k_1$.  In the first subcase, $k_2 t_2 t_1 = k_1 kt$ gives $k_1 = (2,3)k_1 (1,3)$.  This equation has no solutions.  For if $1$ is not in the support of $k_1$, then $k_1$ doesn't move 1 but the right side does.  Since $k_1$ moves 1, 2 and 3 are not in the support of $k_1$ (because $k_1 \ne (1,2),(1,3)$).  So $k_1=(2,3)k_1 (1,3) = k_1 (2,3)(1,3) = k_1 (1,3,2)$, an impossibility.  In the second subcase, we get $k_1 = (1,3,2) k_1 (1,3,2)$.  This equation also has no solutions.  For if the support of $k_1$ is disjoint from $\{1,2,3\}$, then the right side becomes $k_1 (1,2,3)$, which cannot equal $k_1$.  And if the support of $k_1$ overlaps with $\{1,2,3\}$, say $k_1=(1,i)$, then the left side maps $i$ to 1, whereas the right side maps $i$ to 3.

Thus, $k_1 \ne t$ is impossible.

Case 3: Assume $t_1 \ne k$. 
If $t_1 \ne (1,2)$, then since $t_1 \ne t = (2,3)$ and since $t_1 \ne (1,3)$ ($T(S)$ has no triangles), we have that the union of the support of $t_1,k$ and $t$ is at least 4. As before, if $T(S)$ does not have triangles or 4-cycles, then the equation $t_2 k_2 k_1 = t_1 t k$ implies $\{t_2,k_2,k_1\}=\{t_1,t,k\}$. We already have $t_2 \ne t_1$.  As in \cite[p. 71]{Feng:2006}, it can be shown that each of the two subcases $t_2=t$ and $t_2=k$ leads to a contradiction.  Thus, this case is also impossible.
\qed



\bigskip From Lemma~\ref{GodsilRoyle:4cycles}, Lemma~\ref{lemma:feng:normalTS} and Theorem~\ref{thm:6cycle:TS}, it follows that a sufficient condition for normality of $\Cay(S_n,S)$ is that the transposition graph $T(S)$ have girth at least 5:

\begin{Corollary}  Let $S$ be a set of transpositions generating $S_n$ such that the girth of the transposition graph $T(S)$ is at least 5.  Then, the automorphism group of the Cayley graph $\Cay(S_n,S)$ is the semidirect product $R(S_n) \rtimes \Aut(S_n,S)$.
\end{Corollary}

\section{The 4-cycle transposition graph}

Let $S$ be a set of transpositions generating $S_n$.   By Theorem~\ref{theorem:main}, a sufficient condition for normality of $\Cay(S_n,S)$ is that the girth of the transposition graph of $S$ be at least 5.  We now briefly discuss some results concerning the case where the transposition graph $T(S)$ is the 4-cycle graph.  As we mention next, the corresponding Cayley graph $\Cay(S_4,S)$ is not normal; the proofs of these results can be found in \cite{Ganesan:autTS:arXiv:Dec2012}.  

Suppose the transposition graph $T(S)$ is the $n$-cycle graph $(n \ge 4)$, and let $G$ denote the automorphism group of $\Cay(S_n,S)$.  Then, it can be shown  that $G_e$ is the semidirect product $L_e \rtimes D_{2n}$, where $G_e$ is the stabilizer in $G$ of $e$ and $L_e$ is the stabilizer in $G$ of the vertex $e$ and each of its neighbors.  Moreover, $G_e = D_{2n} L_e$, $D_{2n} \cap L_e=1$, and by Biggs \cite[Chapter 16]{Biggs:1993} $L_e$ is a normal subgroup of $G_e$, whence $G_e = L_e \rtimes D_{2n}$.  It can be shown (cf. \cite{Ganesan:autTS:arXiv:Dec2012}) that the vertex neighborhood stabilizer $L_e$ is trivial iff $n \ge 5$.

There is a flaw in the proof by Zhang and Huang \cite{Zhang:Huang:2005} that led to an incorrect result in that paper, and the incorrect result there  was recalled and stated in Feng \cite{Feng:2006} as well; we now recall these statements.   When the transposition graph $T(S)$ is the $n$-cycle graph, the Cayley graph $\Cay(S_n,S)$ is called the modified bubble-sort graph of dimension $n$.  It was claimed in \cite{Zhang:Huang:2005} that when $T(S)$ is an $n$-cycle graph for $n \ge 4$, then 6-cycles satisfying some particular conditions in the Cayley graph $\Cay(S_n,S)$ are unique.  This uniqueness implies that any automorphism of the Cayley graph that fixes a vertex and each of its neighbors is trivial.  It was stated in \cite[p.72]{Feng:2006} that when $T(S)$ is a 4-cycle, by this uniqueness property, the Cayley graph $\Cay(S_4,S)$ is normal and has full automorphism group $R(S_4) \rtimes \Aut(S_4,S) \cong R(S_4) G_e  \cong S_4 \rtimes D_8$, which contains 192 elements.

Actually, it can be shown that when $T(S)$ is the 4-cycle graph, the 6-cycles satisfying certain conditions in the Cayley graph $\Cay(S_4,S)$ are not unique.  In particular, the uniqueness proofs in \cite{Zhang:Huang:2005} are incorrect, and computer simulations have confirmed our results as being correct.   Thus, the vertex neighborhood stabilizer $L_e$ of this Cayley graph need not be trivial.  It can be shown that this stabilizer is the Klein 4-group $\bZ_2 \times \bZ_2$.  With the help of a computer (\cite{GAP4}) it was confirmed that the full automorphism group of $\Cay(S_4,S)$ has $768$ elements and has structure description
$(\bZ_2 \times \bZ_2 \times ((\bZ_2 ^4 \rtimes \bZ_3) \rtimes \bZ_2)) \rtimes \bZ_2$.  The automorphism group of $\Cay(S_4,S)$ can be stated in the form $G=R(S_4)G_e = R(S_4)(L_e \rtimes D_8) \cong R(S_4) (V_4 \rtimes D_8)$.  In this case, $R(S_4)$ is not a normal subgroup of $\Aut(\Cay(S_4,S))$.

More precisely, it can be shown (cf. \cite{Ganesan:autTS:arXiv:Dec2012}) that if $s_i,s_{i+1} \in S$ are two adjacent edges in the transposition graph $T(S) = C_4$, then there are exactly eight distinct 6-cycles in $\Cay(S_4,S)$ that contain $e, s_i, s_{i+1}$ and a vertex at distance 3 from $e$.  The total number of vertices in 6-cycles of $\Cay(S_4,S)$ that contain $e, s_i$ and $s_{i+1}$ and that are at distance 3 from $e$ is exactly 6.  
More generally, if $S$ is a set of transpositions such that $T(S)$ contains a 4-cycle, then there exist $t,k \in S$ such that $tk \ne kt$ and such that there does not exist a unique 6-cycle in $\Gamma$ containing $e, t, k$ and a vertex at distance 3 from $e$.

It can be shown (cf. \cite{Ganesan:autTS:arXiv:Dec2012}) that:
\begin{Proposition}
Let $S$ be a set of transpositions such that the transposition graph of $S$ is the 4-cycle graph. Then the set of automorphisms of the Cayley graph $\Cay(S_4,S)$ that fixes a vertex and each of its neighbors is isomorphic to the Klein 4-group $\mathbb{Z}_2 \times \mathbb{Z}_2$ and hence is nontrivial.  The Cayley graph $\Cay(S_4,S)$ is not normal. 
\end{Proposition}

\bibliographystyle{plain}
\bibliography{refsaut}

\end{document}